\documentstyle[12pt]{article}
\newcommand{\be}{\begin{eqnarray} }
\newcommand{\ee}{\end{eqnarray} }
\newcommand{\beq}{\begin{equation} }
\newcommand{\eeq}{\end{equation} }

\begin{document}
\begin{flushright}
hep-ph/0010298\\
JINR E2-2000-268\\
October 2000

\end{flushright}
\vskip 1cm
\begin{center}
{\Large \bf
Semi-inclusive polarized DIS in terms of Mellin moments.
\\I. Light sea quark polarized distributions
\footnote{Delivered
on October 10 at Physics Workshop "Compass week in Dubna"$\
 $ (JINR, Dubna, 2000) under the title "Polarized sea-quark
 flavor asymmetries and
COMPASS."
}.
\\ $\quad$
}
\\[5mm]
\vskip 1cm
{ \large A.N. Sissakian}\\
{\it Bogolubov  Laboratory of Theoretical
 Physics,
Joint Institute for Nuclear Research,
Dubna, Moscow region 141980, Russia}\\
\vskip 1cm
{ \large O.Yu. Shevchenko}
\footnote{E-mail address: shevch@nusun.jinr.ru}\\
{\it Dzhelepov Laboratory of Nuclear Problems,
Joint Institute for Nuclear Research,
Dubna, Moscow region 141980, Russia}\\
\vskip 1cm
{\large V.N. Samoilov}\\
{\it Scientific Center of Applied Research,
Joint Institute for Nuclear Research,
Dubna, Moscow region 141980, Russia}
\end{center}
\vskip 2cm
\begin{abstract}
In connection with the semi-inclusive polarized DIS,
it is proposed to consider the first Mellin moments $\Delta q$
  of the polarized quark and antiquark densities, instead of
 the respective variables $\delta q(x)$, local  in Bjorken x
 themselves. This gives rise to a very essential simplification
 of the next to leading order (NLO) QCD and, besides, allows
 one to use the respective QCD sum rules.
 An expression for $\Delta\bar{u}-\Delta\bar{d}$ in NLO is
 obtained which is just a simple combination of the directly
 measured asymmetries and of the quantities taken from the
 unpolarized data.
\\
\end{abstract}
\newpage
\vskip 4mm

Investigation of the quark structure of the nucleon
 is one of most important tasks of modern high energy
 physics. In this respect deep inelastic scattering (DIS)
 is of special importance. Thus, the very impressive
 result of the New Muon Collaboration (NMC) experiment was
 obtained in 1991, when the unpolarized structure functions
 of the proton and neutron, $F_2^p(x)$ and $F_2^n(x)$,
 were precisely measured within a wide range of Bjorken's x,
 and, it was established that the integral
$\int_0^1 \frac{dx}{x}[F_2^p(x)-F_2^n(x)]$ does
 not equal $1/3$ (Gottfried sum rule) but has a much
 smaller value $0.235\pm 0.0026$. This means that
 the densities of u and d sea quarks, $\bar{u}(x)$ and
$\bar{d}(x)$, in the proton have different values, and
\be
{ \int_0^1 dx\ [\bar{d}(x)-\bar{u}(x)]=0.147\pm 0.039 \neq 0.\nonumber }
\ee \qquad In polarized DIS, instead of the unpolarized total
$q=q^{\uparrow}+q_{\downarrow}$, sea $\bar{q}$ and valence
$q_V=q-\bar{q}$ quark densities, the set of the respective
polarized quantities $\delta
q(x,Q^2)=q^{\uparrow}(x,Q^2)-q_{\downarrow}(x,Q^2)$,
$\delta\bar{q}(x,Q^2)=\bar{q}^{\uparrow}(x,Q^2)-\bar{q}_
{\downarrow}(x,Q^2)$
and $\delta q_V(x,Q^2)=\delta q(x,Q^2)-\delta\bar{q}(x,Q^2)$ is
the subject of the investigation. So, the question arises:
does the difference between the polarized $u$ and $d$ sea
quark densities $\delta\bar{u}(x,Q^2)-\delta\bar{d}(x,Q^2)$
also differ from zero?
Recently, a series of theoretical papers appeared
 ([1-4]) where it was predicted that the quantity
$\delta\bar{u}(x,Q^2)-\delta\bar{d}(x,Q^2)$
 does not equal zero. However, the
model-dependent results for $\delta\bar{u}(x,Q^2)-\delta\bar{d}(x,Q^2)$
essentially differ each from other in these papers.
So, it is very
desirable to find a reliable way to extract this quantity
directly from experiment data. For this purpose it is not sufficient
 to use
just the inclusive polarized DIS data, and one has to
 investigate  \underline{semi-inclusive} polarized
 DIS processes  like $$\vec{\mu}+\vec{p}(\vec{d})
\rightarrow \mu+h+X.$$

Such processes provide direct access to the individual
 polarized quark and antiquark distributions via
 measurements of the respective \it spin asymmetries\rm.\
 \footnote {Such a kind of measurements were performed by
 SMC and HERMES experiments and are also planned
 by the COMPASS collaboration.}

Unfortunately, the description of semi-inclusive
 DIS processes
turns out to be much more complicated in comparison with the
traditional inclusive polarized DIS. First, the fragmentation
functions are involved, for which no quite reliable
information is available\footnote {For discussion of this subject
 see, for example [5] and references therein. }.
Second (and this is the
most serious problem), the consideration even of the next to
leading (NLO) QCD order turns out to be extremely difficult,
 since it
involves double convolution products. So, to achieve a
reliable description it is very desirable, on the one hand,
to exclude from consideration the fragmentation functions,
whenever possible, and, on the other hand (and this is the
main task), to try to simplify the NLO consideration as much as possible,
without which one can say nothing about the reliability and
 stability of results obtained within the quark-parton model (QPM).

It is well known that within QPM one can completely exclude
the fragmentation functions from the expressions for the
 \underline{valence} quark polarized distributions $\delta q_V$
through experimentally measured asymmetries. To this end,
instead of the usual virtual photon asymmetry $A^h_{\gamma
N}\equiv A_{1N}^h$ (which is expressed in terms of the directly
measured asymmetry $A^h_{exp}=
(n^h_{\uparrow\downarrow}-n^h_{\uparrow\uparrow})/(n^h_
{\uparrow\downarrow}+n^h_{\uparrow\uparrow})$
as $A^h_{1N}=(P_B P_T f D)^{-1}A^h_{exp}$), one has to measure
so called "difference asymmetry" $A^{h^+-h^-}_N$ [6] (see also [5,7])
 which is
expressed in terms of the respective counting rates as
\be
{
A^{h-\bar{h}}_N(x,Q^2;z)=\frac{1}{P_BP_TfD}\frac{(n^{h}_
{\uparrow\downarrow}-n^{\bar{h}}_{\uparrow\downarrow})-(n^{h}_
{\uparrow\uparrow}-n^{\bar{h}}_{\uparrow\uparrow})}{(n^{h}_
{\uparrow\downarrow}-n^{\bar{h}}_{\uparrow\downarrow})+(n^{h}_
{\uparrow\uparrow}-n^{\bar{h}}_{\uparrow\uparrow})},
} \ee where the event densities $n^h_{\uparrow\downarrow
(\uparrow\uparrow)}=dN^h_{\uparrow\downarrow(\uparrow\uparrow)}
/dz$, i.e. $n^h_{\uparrow\downarrow(\uparrow\uparrow)}dz$ are
 the numbers of
events for antiparallel (parallel) orientations of here muon and
 target nuclear  (proton or deutron here) spins
for the hadrons of type h registered in the interval
 $dz$. Coefficients $P_B$ and $P_T$, f and D are
the beam and target polarizations, dilution and depolarization
factors, respectively,(for details on these coefficients see,
 for
example, [8,9] and references therein). Then, the QPM
 expressions
for the difference asymmetries look like (see, for example,
COMPASS project [10], appendix A)
\begin{eqnarray}
A_{p}^{\pi^+-\pi^-}&=&\frac{4\delta u_V-\delta d_V}{4 u_V -d_V}
;\quad A_{n}^{\pi^+-\pi^-}=\frac{4\delta d_V-\delta u_V}
{4d_V-u_V};\nonumber\\
A_{d}^{\pi^+-\pi^-}&=&\frac{\delta u_V+\delta d_V}{u_V+d_V}
;\nonumber\\
A_{p}^{K^+-K^-}&=&\frac{\delta u_V}{u_V};\nonumber\\
A_{d}^{K^+-K^-}&=&A_{d}^{\pi^+-\pi^-},
\end{eqnarray}
i.e., on the one hand, they contain only valence quark
 polarized densities, and, on the other hand, have the remarkable
 property to be free of any fragmentation functions.

All this is very good, but we are interested
here in the sea quark polarized distributions, and, besides,
the main question arises - what will happen with all this beauty
in the next to leading order QCD?

{\bf  We propose} to investigate the \it integral
 \rm quantities,
namely, the first Mellin moments
$M^1(\delta q)\equiv \int_0^1 dx\ [\delta q(x)]\equiv
 \Delta q\ (q=u,d,s,...)$
instead of the local polarized quark densities
$\delta q(x)$ themselves.
This provides very essential \it advantages: \rm

\underline{ First}.

Even if the local quantity has a very small
\footnote{ Notice, however, that the latest theoretical
 paper [4] on this subject predicts that the difference
 between the polarized densities $\delta\bar{u}$ and
$\delta\bar{d}$ should be even more significant than the
difference between the unpolarized sea quark densities:\
$|\delta\bar{u}-\delta\bar{d}|\geq |\bar{u}-\bar{d}|$. }
value at each point x, the integral of this quantity over
the whole range of x-variables may already have quite a
considerable value, and, one can hope that QPM turns out to
be a good  approximation for integral quantities like \be
{\Delta\bar{u}-\Delta\bar{d}\equiv \int^1_0 dx\
[\delta\bar{u}(x)-\delta\bar{d}(x)]. } \ee An argument in
favor of such a hope (for (3)) is the circumstance that all
the model predictions [1-4] have one common feature: the
local quantity $\delta\bar{u}(x)-\delta\bar{d}(x)$ does not
change sign when x varies over its entire range $0\leq
x\leq 1.$

\underline{Second}. \rm\\
 To investigate integral quantities like (3) one can
 use \it QCD sum rules. \rm In particular, one can
 apply such a well established sum rule as \it the
 Bjorken sum rule\rm
 \footnote{Throughout the paper, all the quantities considered in
 NLO are given in the $\overline{\rm MS}$ scheme.
 }
\be
{
\int^1_0 dx[g^p_1-g^n_1]=\frac{1}{6}\
 \frac{g_A}{g_V}(1-\frac{\alpha_s(Q^2)}
{\pi}+O(\alpha_s^2)),
}
\ee
$$ g_A/g_V=1.2537\pm 0.0028 $$
to express the quantity $\Delta\bar{u}-\Delta\bar{d}$
 of interest  via the quantity
$\Delta{u}_V-\Delta{d}_V$ which, in turn, is
 expressed via the measured difference
 asymmetries $A_p^{\pi^+-\pi^-}$ and $A_d^{\pi^+-\pi^-}$.

\underline{ Third}\rm\ (and we consider this \it the most
 important advantage \rm
of the proposed procedure)
\\
Application of the Mellin moments, instead of the
 local quantities themselves, results in a remarkable
 simplification of the NLO QCD consideration of the
 semi-inclusive polarized DIS, that is extremely complicated
 in terms of the local quantities.

\medskip
Thus, let us consider the NLO [11] expression
 for the structure function $g^p_1$
\be
{g^p_1(x,Q^2)=\frac{1}{2}\sum_{q,\bar{q}} e^2_q\left( \delta q
+\frac{\alpha_s(Q^2)}{2\pi}[C_q\otimes
\delta q+C_g\otimes\delta g]\right)(x,Q^2),
}
\ee
where
\be
{
{\Bigl (}C\otimes f{\Bigl )}(x)\equiv \int^1_x\frac{dy}
{y}\ C\left(\frac{x}{y}\right)f(y)
}
\ee
 is the definition of the convolution product. From now on we
 will use the well known remarkable
property of the Mellin n-th moments
\be
{M^n(f)\equiv\int_0^1dx\ x^{n-1}f(x)
}
\ee
to split the convolution product (6) into a simple
product of the Mellin moments of the respective functions:
\be
{
M^n[C\otimes f]\equiv \int^1_0 dx x^{n-1}\int^1_x\frac{dy}
{y}C\left(\frac{x}{y}\right)f(y)=M^n(C)M^n(f).
}
\ee
So, taking the first Mellin moment of Eq. (5) and using
 the expressions for the Mellin moments of the
 respective NLO ($\overline{\rm MS}$) Wilson coefficients
$$
M^1(C_q)=-2,\quad M^1(C_g)=0
$$
one obtains [11] in NLO QCD:

\begin{eqnarray}
M^1[g_1^p] \equiv  \int_0^1 dx \ g_1^p(x,Q^2)=
 \frac{1}{2}\sum_{q,\bar{q}} e^2_q\left(1-\frac{\alpha_s(Q^2)}{\pi}
 \right)
\int_0^1 dx\  \delta q
\end{eqnarray}
and the same for $g_1^n$ with the substitution $u\leftrightarrow
d$. Substituting the last expressions for
$M^1[g_1^p]$ and $M^1[g_1^n]$ into the Bjorken sum rule (4),
 one can
see that the $\alpha_s$ dependent multipliers
$(1-\alpha_s(Q^2)/\pi)$ cancel out precisely in the
left- and right-hand sides, and, one arrives at the
 simple relation
between the polarized densities of sea and valence quarks
\be
{\int_0^1dx\ (\delta\bar{u}-\delta\bar{d})=\frac{1}{2}\
 \frac{g_A}{g_V}-\frac{1}{2}\int_0^1 dx\
(\delta u_V-\delta d_V),
}
\ee
or, in the notation used here,
\be
{
\Delta\bar{u}-\Delta\bar{d}=\frac{1}{2}\ \frac{g_A}{g_V}
-\frac{1}{2}\ (\Delta u_V-\Delta d_V).
}
\ee
\\
Thus,
 the relation between the first Mellin moments of the polarized
 sea and valence quark distributions has a very simple
 form and does not contain $\alpha_s$ dependence at all
 (i.e. is an exact relation at least up to $O(\alpha_s^2)$
 corrections).

 With such a simple relation between
 $\Delta\bar{u}-\Delta\bar{d}$ and $\Delta u_V-\Delta d_V$
  at our disposal,
 {\it the next step} is to establish the relation between
 the Mellin moments
 $\Delta u_V$ and $\Delta d_V$ and the experimentally
 observable difference asymmetries $A_{p(d)}^{\pi^+-\pi^-}$
 in NLO QCD. For this purpose, one can use the following
 relations [5,12,13] for the difference asymmetries
\be
A_N^{h-\bar{h}}(x,Q^2;z)=\frac{g_1^{N/h}-g_1^{N/\bar{h}}}{\tilde{F}_1^{N/h}-
\tilde{F}_1^{N/\bar{h}}}\quad (N=p,n,d),
\ee
where the semi-inclusive analogs of the structure functions
 $g_1^N$ and $F_1^N$, functions $g_1^{N/h}$ and $\tilde{F}_1^{N/h}$,
 $\ $are related to the respective polarized and unpolarized
 semi-inclusive differential cross-sections as follows [12]
\be
\frac{d^3\sigma^h_{N\uparrow\downarrow}}{dxdydz}-\frac
{d^3\sigma^h_{\uparrow\uparrow}}{dxdydz}&=&\frac{4\pi\alpha^2}
{Q^2}\ (2-y)\ g_1^{N/h}(x,z,Q^2),\\
\frac{d^3\sigma^h_N}{dxdydz}&=&\frac{2\pi\alpha^2}{Q^2}\
 \frac{1+(1-y)^2}{y}\ 2\tilde{F}_1^{N/h}(x,z,Q^2),\ee
where
\be
\tilde{F}_1^{N/h}&\equiv& F_1^{N/h}+\frac{1-y}{1+(1-y)^2}
\ F_L^{N/h}.
\ee
The semi-inclusive structure functions $g_1^{p(n)/h}$
 are given in NLO by

\begin{eqnarray}
2g^{p/h}_1&=&\sum_{q,\bar{q}} e_q^2\delta q[1+\otimes
 \frac{\alpha_s}{2\pi}\delta C_{qq}\otimes]D^h_{q}
\nonumber\\&+&(\sum_{q,\bar{q}} e_q^2\delta q)\otimes \frac{\alpha_s}
{2\pi}\delta C_{gq}\otimes D^h_g\nonumber\\&+&\delta g\otimes
 \frac{\alpha_s}{2\pi}\delta C_{qg}\otimes(\sum_{q,\bar{q}} e_q^2
D^h_{q}),
\end{eqnarray}
\be
g_1^{n/h}&=&g_1^{p/h}{\Bigl |}_{u\leftrightarrow d,
\ s\leftrightarrow s},
\ee
where
\be
{
[A\otimes B\otimes C](x,z)\equiv\int\limits_{\cal \quad D}
\int \frac{dx'}{x'}\frac{dz'}{z'} A\left(\frac{x}{x'}
\right)B(x',z')C\left(\frac{z}{z'}\right)
}
\ee
is the double convolution product.
The respective expressions for $2\tilde{F}_1^{p(n)/h}$
 have the same form with the substitution  $\delta q
\rightarrow q$, $\delta C\rightarrow \tilde{C}$.
 The expressions for the Wilson coefficients $\delta C_{qq(qg,gq)}$
 and $\tilde{C}_{qq(qg,gq)}\equiv C^1_{qq(qg,gq)}+2[(1-y)/
 (1+(1-y)^2)]C^L_{qq(qg,gq)}$ can be
 found, for example, in [12], Appendix C.
\\
$\quad\quad$It is remarkable that
due to the properties of the fragmentation functions:
\begin{eqnarray}
D_1&\equiv&D_u^{\pi^+}=D_{\bar{u}}^{\pi^-}=D_{\bar{d}}^{\pi^+}
=D_d^{\pi^-},\nonumber\\
D_2&\equiv&D_d^{\pi^+}=D_{\bar{d}}^{\pi^-}=D_u^{\pi^-}
=D_{\bar{u}}^{\pi^+},
\end{eqnarray}
in the differences
$g_1^{p/\pi^+}-g_1^{p/\pi^-}$ and $
\tilde{F}_1^{p/\pi^+}-\tilde{F}_1^{p/\pi^-}$ (and, therefore, in the
 asymmetries $A_p^{\pi^+-\pi^-}$ and $A_d^{\pi^+-\pi^-}$)
 only the contributions containing the Wilson coefficients $\delta
 C_{qq}$ and $\tilde{C}_{qq}$ survive. However, even then the system of
 double integral equations
\begin{eqnarray}
A_p^{\pi^+-\pi^-}(x,Q^2;z)&=&\frac{(4\delta u_V-\delta d_V)
[1+\otimes \alpha_s/(2\pi)\delta C_{qq}\otimes](D_1
-D_2)}{(4u_V-d_V)[1+\otimes \alpha_s/(2\pi)C_{qq}
\otimes](D_1-D_2)}(x,Q^2;z),\nonumber\\
A_n^{\pi^+-\pi^-}(x,Q^2;z)&=& A_p^{\pi^+-\pi^-}(x,Q^2;z)|_{u_V
\leftrightarrow d_V}
\end{eqnarray}
proposed by E. Christova and E. Leader [5], is extremely
 difficult to solve with respect to the local quantities
 $\delta u_V(x,Q^2)$ and $\delta d_V(x,Q^2)$. Besides, the
 range of integration ${\cal D}$ used in ref. [5] has a very
 complicated form, namely:
\be
{
\frac{x}{x+(1-x)z}\leq x' \leq 1 \ with \ z\leq z'\leq 1,
}
\ee
if $x+(1-x)z\geq 1$, and, additionally,
 range $$x\leq x'\leq x/(x+(1-x))z$$ with $x(1-x')
/(x'(1-x))\leq z' \leq 1$
if $x+(1-x)z\leq 1$. So, one can see that here even application of
the Mellin moments cannot simplify the situation.

Such enormous complification of the convolution
 integral range occurs if one introduces (to take into account
the target fragmentation contributions\footnote{Then, one
 should also add the target fragmentation contributions
 to the right-hand side of (16).} and to exclude the
 cross-section singularity problem at $z_h=0$) a
 new hadron kinematical variable $z=E_h/E_N(1-x)$
 ($\gamma p$ c.m. frame) instead of the usual semi-inclusive
 variable $z_h=(Ph)/(Pq)=(lab. system)\ E_h/E_\gamma$.
 However, both problems compelling us to introduce
  z, instead of $z_h$, can be avoided (see, for example [12,13])
 if one, just to neglect the target fragmentation, applies
 a proper kinematical cut $Z<z_h\leq 1$, i.e. properly
 restricts the kinematical region covered by the final state
 hadrons\footnote{This is just what was done in the HERMES and
 COMPASS experiments, where the applied kinematical cut was
 $z_h>Z=0.2$.}. Then, one can safely use, instead of z,
 the usual variable $z_h$, which at once makes the integration
 range ${\cal D}$ in the double convolution product (18) very
 simple:\footnote{
 Namely the such range was used in the early seminal papers [14]
 (see also [12]).
 }
 $x\leq x'\leq 1,\ z_h\leq z'\leq 1$.
 Note that in applying the kinematical cut it is much more
 convenient to deal with the total numbers of events
 \be
 N^h_{\uparrow\downarrow(\uparrow\uparrow)}(x,Q^2)
 {\Bigl |}_Z=\int_Z^1 dz_h\ n^h_{\uparrow\downarrow(\uparrow\uparrow)}
 (x,Q^2;z_h)
 \ee
 within the entire interval $z\leq z_h\leq 1$ and the
 respective integral difference asymmetries
\footnote{Namely the integral spin symmetries
$ A_{1N}^h=\int_Z^1 dz_h\ g_1^{N/h}\left/\int_Z^1dz_h\
 \tilde{F}_1^{N/h}\right.$
were measured by SMC and HERMES experiments
 (see [8,9] and also [13]).
 }
\begin{eqnarray}A^{h-\bar{h}}_N(x,Q^2){\Bigl |}_Z&=&
\frac{1}{P_BP_TfD}\frac{(N^{h}_
{\uparrow\downarrow}-N^{\bar{h}}_{\uparrow\downarrow})-(N^{h}_
{\uparrow\uparrow}-N^{\bar{h}}_{\uparrow\uparrow})}{(N^{h}_
{\uparrow\downarrow}-N^{\bar{h}}_{\uparrow\downarrow})+(N^{h}_
{\uparrow\uparrow}-N^{\bar{h}}_{\uparrow\uparrow})}{\Bigl |}_Z
=\nonumber\\
&=&\frac{\int_Z^1dz_h(g_1^{N/h}-g_1^{N/\bar{h}})}{\int_Z^1 dz_h(\tilde{F}_1^{N/h}
-\tilde{F}_1^{N/\bar{h}})}\quad (N=p,n,d),\end{eqnarray}
than with the local in $z_h$ quantities $n_{\uparrow\downarrow
(\uparrow\uparrow)}(x,Q^2;z_h)$ and $A_N^{h-\bar{h}}(x,Q^2;z_h)$.
 So, the
 expressions for the proton and deutron integral difference
 asymmetries assume the form\footnote{
 To obtain (25) one uses that the deutron cross section is the
 sum of the proton and neutron cross sections,
 which is valid up to corrections of order $O(\omega_D)$,
 where $\omega_D=0.05\pm0.01$ is the probability to find
 deutron in the $D$-state.}
\begin{equation}
A_p^{\pi^+-\pi^-}(x,Q^2){\Bigl |}_Z=\frac{(4\delta u_V-\delta d_V)
\int_Z^1 dz_h[1+\otimes\frac{\alpha_s}{2\pi}\delta C_{qq}\otimes]
(D_1-D_2)}
{(4u_V-d_V)\int_Z^1 dz_h[1+\otimes\frac{\alpha_s}{2\pi}\tilde{C}_{qq}
\otimes ](D_1-D_2)},
\end{equation}
\be
A_d^{\pi^+-\pi^-}(x,Q^2){\Bigl |}_Z=
\frac{\int_Z^1 dz_h\: [(g_1^{p/\pi^+}-g_1^{p/\pi^-})+(g_1^{n/\pi^+}-g_1^{n/\pi^-})]}
{\int_Z^1 dz_h[(\tilde{F}_1^{p/\pi^+}-\tilde{F}_1^{p/\pi^-})+
(\tilde{F}_1^{n/\pi^+}-\tilde{F}_1^{n/\pi^-})]
}=
\nonumber\\
=\frac{(\delta u_V+\delta d_V)
\int_Z^1 dz_h[1+\otimes\frac{\alpha_s}{2\pi}\delta C_{qq}\otimes]
(D_1-D_2)}
{(u_V+d_V)\int_Z^1 dz_h[1+\otimes\frac{\alpha_s}{2\pi}\tilde{C}_{qq}
\otimes ](D_1-D_2)},
\ee
and the double convolution reads
\be
{ [A\otimes B\otimes
C]=\int_x^1\frac{dx'}{x'}\int_{z_h}^1\frac{dz'}{z'}\
A\left(\frac{x}{x'}\right)B(x',z')C\left(
\frac{z_h}{z'}\right) }. \ee
Now, application of the first Mellin moment to the difference
asymmetries $A_p^{\pi^+-\pi^-}(x,Q^2){|}_Z$ and
 $A_d^{\pi^+-\pi^-}(x,Q^2){|}_Z$,
 given by (24) -- (26), becomes extremely useful and
 allows one to obtain a system of two \it
 purely algebraic\rm\
equations for $\Delta u_V\equiv \int_0^1 dx\ \delta u_V$ and
$\Delta d_V\equiv \int_0^1 dx\ \delta d_V$:
\be
{
(4\Delta u_V-\Delta d_V)(M_1-M_2)={\cal A}_p^{exp},
}
\ee
\be
{
(\Delta u_V+\Delta d_V)(M_1-M_2)={\cal A}_d^{exp},
}
\ee
with the solution
\be
{
\Delta u_V=\frac{1}{5}\frac{{\cal A}_p^{exp}+{\cal A}_
d^{exp}}{M_1-M_2};\quad  \Delta d_V=\frac{1}{5}\frac{4{
\cal A}_d^{exp}-{\cal A}_p^{exp}}{M_1-M_2}.
}
\ee
Here we introduce the notation
\begin{eqnarray}
{\cal A}_p^{exp}&\equiv& \int_0^1 dx\ A_p^{\pi^+-\pi^-}
{\Bigl |}_Z
(4u_V-d_V)\int_Z^1 dz_h[1+\otimes \frac{\alpha_s}{2\pi}
\tilde{C}_{qq}\otimes]
(D_1-D_2),\nonumber \\{\cal A}_d^{exp}&\equiv& \int_0^1 dx\
A_d^{\pi^+-\pi^-}{\Bigl |}_Z(u_V+d_V)\int_Z^1 dz_h[1+\otimes
\frac{\alpha_s}{2\pi}
\tilde{C}_{qq}\otimes](D_1-D_2),\end{eqnarray}
and
\begin{eqnarray}
M_1&\equiv& M_u^{\pi^+}=M_{\bar{u}}^{\pi^+}=M_{\bar{d}}^{\pi^+}
=M_d^{\pi^-},
\nonumber\\
M_2&\equiv& M_d^{\pi^+}=M_{\bar{d}}^{\pi^-}=M_u^{\pi^-}
=M_{\bar{u}}^{\pi^+},
\end{eqnarray}
where
\be
{
M_q^h(Q^2)\equiv \int_Z^1 dz_h\left[D_q^h(z_h,Q^2)+
\frac{\alpha_s}{2\pi}\int_{z_h}^1
\frac{dz'}{z'}\ \Delta C_{qq}(z')D_q^h(\frac{z_h}{z'},Q^2)\right]
}
\ee
with the coefficient
\be
{
\Delta C_{qq}(z)\equiv \int_0^1 dx\ \delta C_{qq}(x,z),
}
\ee
that is given in Appendix.
Thus, using the relation (11) between $\Delta u-\Delta d$
 and $\Delta
u_V-\Delta d_V$ one gets,
eventually, a simple expression for
$\Delta\bar{u}-\Delta\bar{d}\equiv\int_0^1dx\
(\delta\bar{u}(x,Q^2)-\delta\bar{d}(x,Q^2))$ in terms of
experimentally measured quantities, that is valid in NLO QCD :
\be
{
\Delta\bar{u}-\Delta\bar{d}=\frac{1}{2}\ \frac{g_A}{g_V}-\frac{2{\cal A}
_p^{exp}-3{\cal A}_d^{exp}}{10(M_1-M_2)}.
}
\ee
It is easy to see that all the quantities present in the
 right-hand side, with the exception of the  two difference
 asymmetries
 $A_p^{\pi^+-\pi^-}{|}_Z$ and $A_d^{\pi^+-\pi^-}
 {|}_Z$ (entering into ${
\cal A}_p^{exp}$ and ${\cal A}_d^{exp}$, respectively) can
 be extracted from \it unpolarized\rm \footnote{With the
 standard and well established assumption that the
 fragmentation functions do not depend on the spin.
 Then, the unpolarized fragmentation functions D can be
 taken either from independent measurements of $e^+e^-$ -
 annihilation into hadrons [15] or in hadron production in
 unpolarized DIS [16]} semi - inclusive data
 and can, thus, be considered here as a known input.
 So, the only quantities that have to be measured in {\it
 polarized}  semi-inclusive DIS are the difference
 asymmetries $A_p^{\pi^+-\pi^-}{|}_Z$ and
 $A_d^{\pi^+-\pi^-}{|}_Z$
 which, in turn, are just simple combinations of the
 directly measured counting rates.

In conclusion, we would like to stress that application
 of the
Mellin moments, instead of the local polarized densities,
 happens to be very
fruitful not only in the case of light u- and d-quarks, but also for
investigation of polarized strangeness in the nucleon (a
paper is now in preparation). Besides, we also plan to
apply this procedure to the transverse asymmetries in the
nearest future.

 At present, a proposal for measurement of
 $\Delta\bar{ u}-\Delta\bar{d}$ , based on the above
 described procedure, is being prepared for the experiment
 COMPASS in collaboration with the group of INFN -- sezione
 di Torino and of Dipartimento di fisica generale "A.Avogadro"
 of the Torino  University.
\\\\
\ \ \ \ \  The authors are grateful to R.~Bertini, M.~P.~Bussa,
 O.Yu.~Denisov, A.V. Efremov, O.N. Ivanov, V.G. Kadyshevsky,
 V. Kallies, N.I. Kochelev, A.M.~Kotzinian, E.A.~Kuraev,
 A.~Maggiora, G.~Piragino, G.~Pontecorvo for
 fruitful discussions, and one of us (O.Yu.~Shevchenko) for the
 hospitality and friendly atmosphere he met with in Torino. We
 also wish to thank F.~Bradamante, A.E. Dorokhov, M.G.~Sapozhnikov and
 I.A. Savin for interest in this work.
\begin{center}
{\bf Appendix: Mellin moments of polarized semi-inclusive DIS
 Wilson coefficients.}
 \end{center}
The NLO ($\overline{MS}$) coefficient  $\delta C_{qq}$ has the form
(see [12], Appendix C)
\be\delta C_{qq}=C_{qq}^1-2C_F(1-x)(1-z)\quad(C_F=4/3),\ee
where

\begin{eqnarray}
C^1_{qq}&=&C_F\left\{ -8\delta(1-x)\delta(1-z)+\delta(1-x){\Bigl [}\tilde{P}_{qq}(z)\ln\frac{Q^2}{M^2_F}+L_1(z)+L_2(z)\right.\nonumber\\
&+&(1-z){\Bigl ]}+\delta(1-z)\left[\tilde{P}_{qq}
(x)\ln\frac{Q^2}{M^2}+L_1(x)-L_2(x)+(1-x)\right]+\nonumber\\
&+&2\frac{1}{(1-x)_+}\frac{1}{(1-z)_+}-\left.\frac{1+z}
{(1-x)_+}-\frac{1+x}{(1-z)_+}+2(1+xz)\right\},
\end{eqnarray}
\be
\tilde{P}_{qq}(z)=\frac{1+z^2}{(1-z)_+}+
\frac{3}{2}\delta(1-z),\nonumber\\
L_1(z)=(1+z^2)\left(\frac{\ln(1-z)}{1-z}\right)_+,
\quad L_2(z)=\frac{1+z^2}{1-z}\ln z,
\ee
and the "+" prescription is given by
\be
\int_0^1dz\ f(z)(g(z))_+\equiv\int_0^1dz\ [f(z)-f(1)]\ g(z).
\ee
Simple calculation of the n-th moment $M^n({\delta C_{qq}})
$ gives
\be
M_n(\delta C_{qq})=
C_F\left\{\delta(1-z)\left [-8
+\frac{3}{2}\ln\frac{Q^2}{M_F^2}+\ln\frac{Q^2}{M^2}
\left(\frac{3}{2}-2\gamma
-\frac{1}{n}-\frac{1}{1+n}-2\Psi(n)\right)\right.\right.
\nonumber\\
+\frac{1}{6}\left(6\gamma^{2}+3\left(\frac{1}{n^2}+
\frac{1}{(1+n)^2}\right)\right.
+6\gamma\left(\frac{1}{n}+
\frac{1}{1+n}\right)
+\pi^2+12\gamma\Psi(n)+3\Psi^2(n)
\nonumber\\
+\left.\left.3\Psi^2(n+2)-6\frac{d\Psi(n)}{dn}\right)
+\zeta(2,n)+\zeta(2,2+n)+\frac{1}{n(n+1)}\right ]\nonumber\\
-\frac{2}{(1-z)_+}[\gamma+\Psi(n)]
+(1+z)[\gamma+\Psi(n)]\nonumber\\
-\frac{1}{(1-z)_+}
\left(\frac{1}{n}+\frac{1}{1+n}\right)+2\left(\frac{1}{n}
+\frac{z}{1+n}\right)
\nonumber\\
+\left.\tilde{P}_{qq}(z)
\ln\frac{Q^2}{M_F^2}+L_1(z)+L_2(z)
+(1-z)\frac{n^2-3n-2}
{n(n+1)}\right \},\nonumber
\ee
where $\Psi(z)=\Gamma'(z)/\Gamma(z)$; $\gamma\simeq$
 0.577216 is the Euler constant.

For example, the first two moments are
\be M_1\left(\delta C_{qq}\right)\equiv\Delta C_{qq}
=C_F\left[1+2z-\frac{3}{2}\frac{1}{(1-z)_+}+\delta(1-z)
\left(-7+\frac{\pi^2}{3}+\frac{3}{2}\ln \frac{Q^2}{M_F^2}
\right)\right.+\nonumber\\
+\left.\tilde{P}_{qq}(z)\ln\frac{Q^2}{M_F^2}+L_1(z)
+L_2(z)\right];
\ee
\be
M_2\left(\delta C_{qq}\right)=C_F\left[\frac{5}{3}+2z
-\frac{17}{6}\frac{1}{(1-z)_+}+\frac{1}{6}\delta(1-z){\Bigl(}-41+2
\pi^2-\right.
\nonumber\\
8\ln\frac{Q^2}{M^2}+9\frac{Q^2}{M_F^2}{\Bigl )}
+\left.\tilde{P}_{qq}(z)\ln\frac{Q^2}{M_F^2}+L_1(z)+L_2(z)
+\frac{2}{3}(1-z)\right].
\ee


\begin{thebibliography}{99}
\bibitem{ref2}
A.E. Dorokhov, N.I. Kochelev, Phys. Lett. {\bf B 304}
(1993) 167;\\ A.E. Dorokhov, N.I. Kochelev and Yu.A. Zubov,
 Int. J. Mod. Phys. {\bf A 8} (1993) 603.
\bibitem{ref3} D.I. Diakonov, V.Yu. Petrov, P.V. Pobylitsa,
 M.V. Polyakov and C. Weiss, Nucl. Phys. {\bf B 480} (1996)
 341; Phys. Rev. {\bf D 56} (1997) 4069.
\bibitem{ref1} R.J. Fries and A. Sch{\"a}fer, Phys. Lett.
 {\bf B 443} (1998) 40.
\bibitem{ref4}B. Dressler, K. Goeke, M.V. Polyakov and C. Weiss,
 Eur. Phys. J. {\bf C 14} (2000) 147.

\bibitem{ref5}E. Christova and E. Leader, hep-ph/0007303.

\bibitem{ref6} L.L. Frankfurt, M.I. Strikman, L. Mankiewicz,
 A. Sch{\"a}fer, E. Rondio, A. Sandacz and V. Papavassiliou,
 Phys. Lett. {\bf B 230} (1989) 141;\\
 F.E. Close, R.G. Milner, Phys. Rev. {\bf D 44} (1991) 3691.

\bibitem{ref7} D. de Florian, L.N.~Epele, H.~Fanchiotti,
 C.A.~Garcia~Canal, S.~Joffily, R.~Sassot, Phys. Lett.
 {\bf B 389} (1996) 358.



\bibitem{ref8}SMC collaboration ( D. Adams et al.), Phys. Rev.
 {\bf D 56} (1997) 5330;\\
 SMC collaboration (B.~Adeva et al.), Phys. Lett. {\bf B 369} (1996) 93;
 Phys. Lett. {\bf B 420} (1998) 180.

\bibitem{ref9}
HERMES collaboration (K.~Ackerstaff et al.), Phys. Lett.
 {\bf B 464} (1999) 123.
\bibitem{ref10}COMPASS collaboration (G. Baum et al.),
"COMPASS: A proposal for a common muon and proton apparatus
 for structure and spectroscopy", CERN-SPSLC-96-14 (1996).

\bibitem{ref11} M. Gl{\"u}ck, E. Reya, M. Stratmann and W.
 Vogelsang, Phys. Rev. {\bf D 53} (1996) 4775.



\bibitem{ref12} D. de Florian, M.~Stratmann and W.~Vogelsang,
Phys. Rev. {\bf D 57} (1998) 5811.


\bibitem{ref13} D.~de~Florian and R.~Sassot, hep-ph/0007068.

\bibitem{ref14} G. Altarelli, R.K. Ellis, G. Martinelli, So-Yong Pi,
Nucl. Phys. {\bf B 160} (1979) 301;\\
W. Furmanski, R. Petronzio, Z. Phys. {\bf C 11} (1982) 293.

\bibitem{ref15} J. Binnewies, B.A. Kniehl, G. Kramer,
 Z. Phys. {\bf C 65} (1995) 471.
\bibitem{ref16} European Muon Collaboration
 ( M. Arneodo et al. ), Nucl. Phys. {\bf B 321} (1989) 541.


\end{thebibliography}
\end{document}